# Mitigating the Impact of Distributed Generations on Relay Coordination Using Fault Current Limiters

Meisam Ansari [1*], Mostafa Ansari [2]

[1] *Department of Electrical and Computer Engineering, Southern Illinois University of Carbondale, USA*
[2] *Department of Electrical and Computer Engineering, Isfahan University of Technology, Iran*

| Keywords | Abstract |
|---|---|
| Distributed Generation, Fault Current Limiters, Overcurrent Relay Protection. | The use of distributed generation resources, in addition to considerable benefits, causes some problems in the power system. One of the most critical problems in the case of disruption is increasing short-circuit current level in grids, which leads to change the protection devices settings in the downstream and upstream grid. By using fault current limiters (FCL), short-circuit currents in grids with distributed generation can be reduced to acceptable levels, so there is no needed to change the protection relays settings of the downstream grid (including distributed generations). However, by locating the FCL in the tie-feeder, the downstream grid is not more effective than the upstream grid and thus its reliability indices also will be changed. Therefore, this paper shows that by locating the unidirectional fault current limiter (UFCL) in the tie-feeder, the necessity of changing in the relay protection settings of upstream grids is prevented. In this paper, the proposed method is implemented, and its efficiency is reported in six scenarios. |

## 1. Introduction

The development of urbanization and industrialization of societies has increased electrical energy demand. Despite developments in power grids, many development programs have been limited because of economic reasons and environmental viewpoints [1], such as constructing new transmission lines and larger power plants, etc.. Under these conditions, renewable energy resources connected to the distribution grids are increasingly developed.

Distributed generation units based on renewable energy sources such as photovoltaic systems, wind turbines will play a vital role in reducing emissions of greenhouse gases [1]. The use of distributed generations (DGs), despite many benefits, can significantly impact power flow, harmonic generation, and increasing short-circuit current in the radial distribution grids [2-6].

Generally, the use of DGs in the grid affects the amplitude and direction of the fault current. It disrupts coordination between protection devices installed in the distribution and sub-transmission system, leading to the false trip of feeders, blinded protection, unintended islanding, and increasing the short-circuit level [7-9].

Therefore, studies have been done in this area in order to solve miss-coordination between protection relays in presence distributed generation [7, 10, 11].

One of the easiest ways to maintain coordination of protection devices is disconnecting DG from the grid when the fault occurs. Power electronic devices such as gate turn off thyristor (GTO), are used instead of circuit breakers for disconnecting DG's during the fault [12]. The reasons of using GTO are high switching speed, high capacity of power flow, high-voltage level and high fault current capacity [13, 14]. The most crucial drawback of disconnecting DG's to maintain coordination of relays is that re-connecting those to the network may be confronted with the problem of synchronization. In [15], with distributed generation units, coordination of grid protection devices such as fuses, high current relays, re-closer be reviewed again. In cases where coordination between protection devices is missed, protection devices settings and, if needed, protection devices will be changed to finally coordinating protection devices with the presence of distributed generations also be established [16, 17].

Another method that can be used for maintaining relay coordination is adaptive protection. Based on the type of DERs and configuration of the network, the setting of protective devices will be changed accordingly to coordinate relays. Based on the system's different topology, the required changes for a relay are sent to the relays, updating the settings through communication links [18-21]. One of the comprehensive methods for adaptive protection is using PMU's data [22]. To achieve a 100% observable system, the easiest approach is to install the PMUs at every bus of the system. However, the cost of installation will be too much. For this reason, different scenarios are considered, such as the impact of zero injection bus, line and PMU outage, etc. to completely address the PMU allocation problem which is discussed thoroughly in [23].

Another way to deal with increasing short-circuit levels of the grid in the presence of DG is to limit the fault current. FCL is used to reduce the short-circuit level in the grid. The performance of an ideal FCL is in the way that before the fault occurs, FCL has no impact on the grid and load flow, and when a fault occurs, by using voltage analysis and harmonic analysis detects the fault current and enter a large-impedance into the grid in series. By installing the FCL, there is no needed to change relays setting and equipment such as circuit-breakers, measuring instruments since the fault current level is not changed [24-26].

---

[*] Corresponding Author: Meisam Ansari
E-mail address: meisam.ansari@siu.edu

Due to the importance of this subject, many studies have been done so far. [27] has studied different locations to install FCL and showed that the best location to install the FCL to mitigate the effect of DG on protection is the DG's branches. So, installing FCL in series with DG can limit fault currents in different situations for all operating conditions of the DG to maintain protection coordination [28].

Loss of coordination between protection relays of downstream and upstream grids is one of the problems that can occur. To overcome this problem, unidirectional fault current limiters (UFCL) are designed and recommended for the microgrid.

UFCL shows little resistance in normal conditions and while a fault occurs in the downstream and shows high resistance when a fault occurs in the upstream grid.

Given the above description, using and adjusting UFCL in smart grids has been rarely investigated in previous studies. For this purpose, in this paper, the application of FCL in smart grid consists of two DG, preventing making changes in protection settings and improving grid reliability is evaluated [29].

The paper is organized as follows: the DG's impact on overcurrent relays is summarized in Section II. Types of and FCL its operation behaviour is explained in Section III. Section IV presents simulation results and analysis the impact of UFCL on power system protection and relay's coordination. Conclusions are summarized in Section V.

## 2. DG Impact on Overcurrent Relays and Coordination Constraints

Operation time of overcurrent relays is a function of the current flowing through the relays and relay settings. It can be written as Eq. (1):

$$\frac{t}{T.D.S} = B + \frac{A}{-1+M^C} \quad , \quad \frac{I_{SC}}{I_P} = M \tag{1}$$

Where $t$ is the operating time of relay, T.D.S is the time dial setting, Where $t$ is the operating time of relay, T.D.S is the time dial setting, $I_p$ is the relay's pickup current, and $I_{sc}$ is the fault current flowing through relay. Coefficients A, B, and C are constant coefficients, dependent on the type of relay.

To coordinate overcurrent relays on the grid, the time interval between main and backup relays must be in the range of (0.3-0.6) second, depends on the type of equipment and grid structure [19, 30].

As mentioned, in the presence of DG's in distribution grids, short-circuit current is increased. Then, it will cause serious problems in coordination of existing relays, and the prior setting will not be valid. Adjusting and coordinating the overcurrent relays (including pickup currents, T.D.S, selecting time intervals to coordinate) must be done [31].

Figure 1 shows that by installing DG's, short-circuit current is increased, and. As Figure 1 implies, the time interval between two relays (main and backup relay) is decreased so, the backup relay may operate earlier than main relay, unnecessarily, so some parts of the grid may experience power outage without any problem.

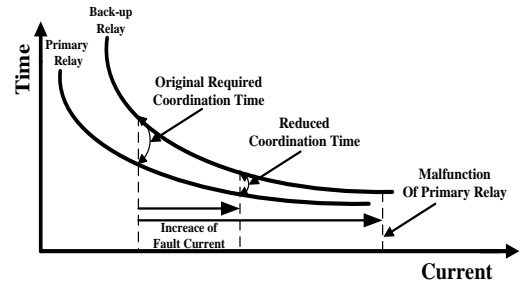

**Figure 1.** Effect of DGs on short-circuit level [1].

## 3. Using (FCL) to Maintain Protection Coordination

FCL is used to reduce the short-circuit level in grid. The performance of an ideal FCL is in the way that in a normal state, FCL has no impact on the grid and load flow, and when a fault occurs, by using voltage analysis and harmonic analysis detects the fault current and enter a large-impedance into the grid in series. This impedance is caused to limit the fault current [32-34].

Many parameters are involved in determining the impedance of the FCL. However, generally, when the fault occurs, the impedance of F.C.L. is chosen to maintain the level of short circuit with or without D.G.s constant. In this way, the setting of protection devices will not be changed, and coordination will be maintained, as shown in Figure 2.

Therefore, the presence of FCL reduces current flowing by DG and the coordination time interval will be maintained.

Fault Current limiters can be divided into passive limiters, static limiters, and hybrid limiters [35, 36].

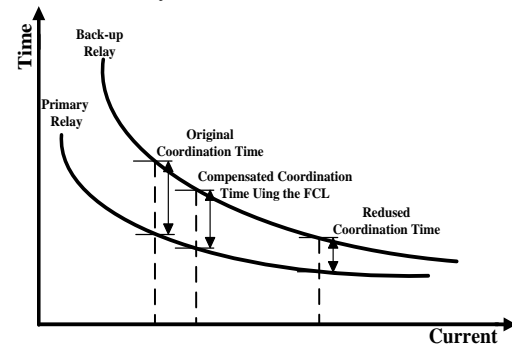

**Figure 2.** Using FCL to maintain relay coordination

Passive limiters do not require external triggers. One of the simplest passive limiters is inductive limiters. Since the current cannot be changed instantaneously in the inductor; the fault current is not decreased significantly at the time of the incident. These limiters. Although having low maintenance costs, these limiters have huge volumes and challenging to move [34].

Another type of passive limiter is the superconducting current limiter (SFCL). SFCLs operate based on this rule that, during the fault, the high current leads to an increase in the temperature and impedance of the superconducting material. this temperature and impedance increase rapidly . The superconducting limiters have a fast response time and do not cause any voltage drop in the steady-state. Nevertheless, its cooling technology is still in its early stages, which leads to consistent failure in its operation.

Another method for decreasing the fault current is using Continuously Variable Series Reactor (CVSR). CVSR is a series reactor that can be used as a fault current limiters in the power system. Its reactance can be changed by applying different DC biases [37]. CVSR can add additional impedance into the ac circuit to decrease fault currents. Its reactance is maximum when the core completely works in a linear region and reaches the minimum value when the core is fully saturated [38].

On the other hand, when a fault occurs in the downstream grid, limiting the short-circuit current by FCL can negatively impact the flexibility and reliability of the downstream grid [28].

### 3.1. The Use of Unidirectional Fault Current Limiter (UFCL)

UFCL is one of the new types of FCL that locates in the grid in series. So that, when a fault occurs in the downstream grid of the UFCL location, the current from the upstream grid that feeds the downstream grid, isn't limited by the UFCL. However, if a fault occurs in the upstream grid of the UFCL location, the current is limited by UFCL, and then, no current is flowed from the downstream grid to the fault location (at the upstream grid). As mentioned, short-circuit current can be detected by using different algorithms [39].

It should be noted that in this paper, UFCL is considered as a passive (resistance) element, and its value determined in the way while a short-circuit occurs in the upstream grid and short circuit level restores its value before the presence of DG in the downstream grid

### 4. Numerical Studies and Results

In this section, at first, the load flow without the presence of DG's and UFCL on the studied grid in PSCAD are calculated Then, the relays are coordinated. After that, fault analysis in the presence of DG is done and the time interval between relays operation is calculated. By inserting the UFCL, its impact on the relays coordination is studied in six scenarios.

### 4.1. Using the Studied Grid

The studied grid is shown in Figure 3, includes an infinite grid, 7buses (6 buses of 20 kV and 1 bus of 400 V) 6 feeders, 3 transformers and 6 overcurrent relays. The studied grid is fed by an infinite grid and adding 2 same DG that are located at buses of DG1 and DG2 [18].

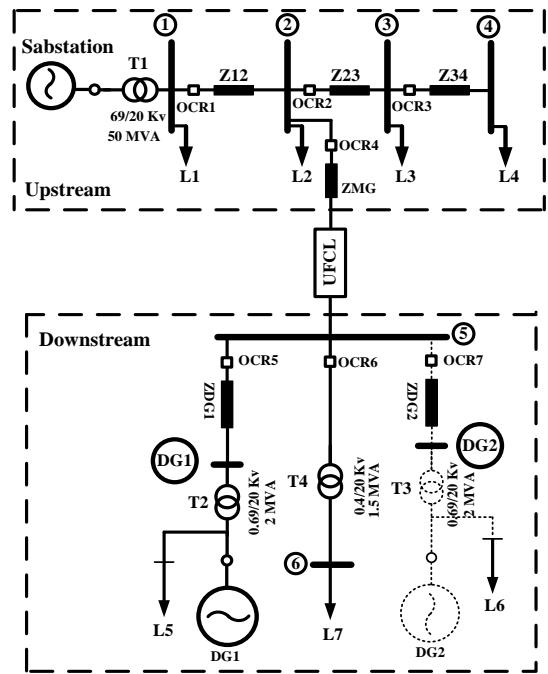

**Figure 3.** The studied grid.

**Table 1.** Grid parameters

| Grid's components | Resistance ($\Omega$) | Reactance ($\Omega$) |
|---|---|---|
| Main (Infinite) Grid | 0.0134 | 0.292 |
| Transformer 1 | 0.0256 | 1.639 |
| Transformers (2&3) | 2.2 | 11.79 |
| Transformer (4) | 2.66 | 17.12 |
| Distributed Generation Unit (DG1&DG2) | 0.0085 | 0.057 |
| L1- L4 | 9.4 | 3.48 |
| L5&L6 | 1.14 | 0.37 |
| L7 | 0.873 | 0.22 |

### 4.2. Calculate the Load Flow and Set the Overcurrent Relays

Load flow is calculated in studied network, in the absence of DGs and based on current of each lines then relay's setting will be calculated (T.D.S, and pickup current). The simulation results are as follows in table2 (numbers are in RMS):

**Table 2.** Relay's setting

| TDS (S) | $I_P$ (A) | Load Flow (A) | Number of Relay |
|---|---|---|---|
| 0.6 | 610 | 558 | Relay 1 |
| 0.4 | 380 | 335 | Relay 2 |
| 0.3 | 260 | 170 | Relay 3 |
| 0.5 | 70 | 48 | Relay 4 |
| 5.5 | 40 | 26.6 | Relay 5 |
| 0.1 | 28 | 18.3 | Relay 6 |

### 4.3. Fault Analysis on the Studied Grid without Distributed Generations

According to the studied grid without DG, a three-phase fault is applied to buses 3,4,6 and the DG's buses, results are showed in Table (3).

**Table 3.** Fault currents on the studied grid

| Relay Flow (A) | Number of Relay (Main and Backup) | Fault Location |
|---|---|---|
| 981.81 | Relay2 (Main) | Bus 3 |
| 984.72 | Relay1 (Backup) | |
| 684.9 | Relay3 (Main) | Bus 4 |
| 684.9 | Relay2 (Backup) | |
| 287.8 | Relay6 (Main) | Bus 6 |
| 313.95 | Relay4 (Backup) | |
| 1091.77 | Relay5 (Main) | DG's bus |
| 1091.77 | Relay4 (Backup) | |

### 4.4. Short-Circuit in the Presence of DG1 and Absence of UFCL (Scenario 1)

In presence of DG1, a three-phase short-circuit is applied on buses 3,4 and 6, and DG's bus. Current passes through each relays and operating time of relays are shown in Table (4). It is important to mention that acceptable time interval between 2 main and backup relays is considered as (0.3-0.6 (s)). It must be mention that in this scenario it is assumed that the type of DG1 is synchronous generators. The relays operation time and the current passes through each relays are shown in table 4.

As shown, while a three phase fault occurs in bus 4, the time interval between relays 2 and 3 is less than 0.3 (s) as well as for relays 1 and 2 is more than 0.6 (s) for three phase fault on bus 3 can disrupt the coordination between relays in the grid.

### 4.5. Short-Circuit in the Presence of DG1 and UFCL

In this scenario, it's tried to solve the problem in the coordination of overcurrent relay that recently showed when a three- phase fault occurs. The relays coordination will be maintained as proposed by locating a passive UFCL in the tie-feeder of the upstream and downstream grid (microgrid).

**Table 4.** Relay's operating time for three phase fault on different buses in presence of DG1 and without UFCL

| Interval (s) | Performance Time (s) | Relay Flow(A) | Number of Relay (Main and Backup) | Fault Location |
|---|---|---|---|---|
| **0.7*** | 0.39 | 1066.52 | Relay2 (Main) | Bus 3 |
| | 1.09 | 938.68 | Relay1 (backup) | |
| **0.28*** | 0. 21 | 727.2 | Relay3 (Main) | Bus 4 |
| | 0.49 | 727.2 | Relay2 (backup) | |
| 0.313 | 0.029 | 308.72 | Relay6 (Main) | Bus 6 |
| | 0.342 | 260.2 | Relay4 (backup) | |
| 0.3691 | 0.4521 | 1091.7 | Relay5 (Main) | DG's bus |
| | 0.083 | 1091.7 | Relay4 (backup) | |

**Table 5.** Relay's operating time for three phase fault on different buses in presence of DG1 and UFCL

| Interval (s) | Performance Time (s) | Relay Flow(A) | Number of Relay (Main and Backup) | Fault Location |
|---|---|---|---|---|
| 0.55 | 0.46 | 996.9 | Relay2 (Main) | Bus 3 |
| | 1.01 | 982.1 | Relay1 (backup) | |
| 0.341 | 0/221 | 686.1 | Relay3 (Main) | Bus 4 |
| | 0.562 | 686.1 | Relay2 (backup) | |
| 0.313 | 0/029 | 308.72 | Relay6 (Main) | Bus 6 |
| | 0/342 | 260.2 | Relay4 (backup) | |
| 0.3691 | 0.4521 | 1091.77 | Relay5 (Main) | DG's bus |
| | 0.083 | 1091.77 | Relay4 (backup) | |

. The coordination time interval between each main and backup pairs is acceptable. Simulation results are shown in Table (5) (in this section, the resistance of UFCL is considered as 184 Ω). So, it is concluded that UFCL provides an acceptable the time interval between relays (2 and3) as well as relays (2and1).

As the results of this simulation show, the operation time intervals between overcurrent relays (1and2), and relays (2and3) when a fault occurs on bus 3, and 4 respectively by selecting an appropriate value of UFCL and using a passive UFCL are reached from unacceptable values (0.7 and 0.28(s)) to acceptable values (0.55 and 0.341 (s)), respectively and then, the protection reliability of the grid in the presence of DG will be maintained. It must be noted that according to the simulation results, during the fault in the downstream grid, there is no difference between the fault currents flow through relays (4,5 and 6) in the presence of UFCL, and absence of UFCL. So, the relays operation time in these case will not be different that much.

### 4.6. Short-Circuit in the Presence of DG1 and DG2 and in the Absence of UFCL (scenario 3)

In addition to DG1, DG2 is also added to the studied grid and like scenarios (1 and 2), three-phase faults occur at different buses. Simulation results are shown in Table (6).

**Table 6.** Simulation results while the occurrence of short-circuit in the presence of DG1, DG2, and UFCL

| Interval (s) | Performance Time (s) | Relay Flow (A) | Number of Relay (Main and Backup) | Fault Location |
|---|---|---|---|---|
| **0.89*** | 0.29 | 1124.2 | Relay2 (Main) | Bus 3 |
|  | 1.18 | 897.3 | Relay1 (backup) |  |
| **0.265*** | 0.28 | 752.3 | Relay3 (Main) | Bus 4 |
|  | 0.473 | 752.3 | Relay2 (backup) |  |
| 0.334 | 0.024 | 318.2 | Relay6 (Main) | Bus 6 |
|  | 0.358 | 236.17 | Relay4 (backup) |  |
| 0.33 | 0.413 | 1376.02 | Relay5 (Main) | DG's bus |
|  | 0.083 | 1084.7 | Relay4 (backup) |  |

**Table 7.** Simulation results while the occurrence of short-circuit in the presence of DG1 and DG2 and in the absence of UFCL

| Interval (s) | Performance Time (s) | Relay Flow (A) | Number of Relay (Main and Backup) | Fault Location |
|---|---|---|---|---|
| 0.53 | 0.49 | 995.6 | Relay2 (Main) | Bus 3 |
|  | 1.02 | 987.8 | Relay1 (backup) |  |
| 0.3278 | 0.235 | 697.2 | Relay3 (Main) | Bus 4 |
|  | 0.5628 | 697.2 | Relay2 (backup) |  |
| 0.334 | 0.024 | 318.2 | Relay6 (Main) | Bus 6 |
|  | 0.358 | 236.17 | Relay4 (backup) |  |
| 0.33 | 0.413 | 1376.02 | Relay5 (Main) | DG's bus |
|  | 0.083 | 1084.7 | Relay4 (backup) |  |

When a fault occurs at buses (3 and 4), it is seen that the operation time interval between relays 1 and 2, and relays 2 and 3 (short-circuit in bus 4) are 0.89 (s) and 0.265 (s), respectively. These are not in a proper range, and therefore, loss of coordination between relays is possible.

*4.7. Short-Circuit in the Presence of DG1, DG2, and UFCL (scenario 4)*

In the grid with DG1 and DG2, by entering the UFCL and, like pervious scenarios, apply three phase faults on different buses. The results can be seen in Table (7). (In this section, the resistance of UFCL is considered 196 Ω to maintain the coordination between overcurrent relays).

It can be seen in Tables 6 and 7 that operation time intervals between overcurrent relays 1 and 2 (main and backup relays) when the fault occurs on bus 3 and overcurrent relays (2 and 3) (main and backup relays) when the fault occurs on bus 4, are reached from 0.89 and 0.265 (s) to 0.53 and 0.3278 (s), respectively. Therefore, the protection reliability of the grid with 2 DG will be maintained by using UFCL.

*4.8. H. Short-circuit in the presence of induction Generator DG1 and in the absence and presence of UFCL (scenario 5 and 6)*

In this case, instead of a synchronous generator, an asynchronous generator is connected to the grid to investigate the effect of these generators on the coordination of relays Without using UFCL. It must be mention that when the fault occurs these types of generator does not have a much difference in comparison to the synchronous generators. So, at first, in the presence of DG1, the three-phase faults apply at different buses without UFCL as table 8 depicts that the operation time interval between relay 1 and 2, and relay 2 and 3 are 0.836 and 0.279, respectively.

scenarios, the operation time interval between relays is calculated. As shown in table 9, the Coordination time interval between relay 1 and 2 and relay 2 and 3 are restored to 0.574 0.348, respectively.

Moreover, coordination between other relays are preserved. Other types of DGs are inverter-based DGs. The inverter-based DGs are DGs that are connected to the power grid through the inverters. Some of these DGs are equipped with LVRT is the capability of a DG's to remain connected to the network during faults, providing voltage, for a time duration that depends on the voltage drop at the point of common coupling (PCC).

Since LVRT requirements keep DERs connected to the network during faults, the extended short-circuit contribution of DG, imposed by the LVRT requirements, can result in significant protection issues such as blinding of protection and directionality issues.

So, the line protection scheme might operate before a downstream DG-unit disconnected due to the LVRT requirements. As the DG unit may continue its operation, the isolated line may keep energized by the DG unit causes unintentional islanding.

In other words, the quick relay operating may lead to an unintentional islanding situation, where the isolated feeder keeps being energized by the DGs.

**Table 8.** Relay's operating time for three phase fault on different buses in presence of DG1 and without UFCL

| Interval (s) | Performance Time (s) | Relay Flow (A) | Number of Relay (Main and Backup) | Fault Location |
|---|---|---|---|---|
| **0.836*** | 0.374 | 1073.3 | Relay2 (Main) | Bus 3 |
|  | 1.21 | 912.8 | Relay1 (backup) |  |
| **0.279*** | 0.191 | 730 | Relay3 (Main) | Bus 4 |
|  | 0.47 | 730 | Relay2 (backup) |  |
| 0.301 | 0.024 | 311.63 | Relay6 (Main) | Bus 6 |
|  | 0.335 | 251.2 | Relay4 (backup) |  |
| 0.365 | 0.444 | 1086.3 | Relay5 (Main) | DG's bus |
|  | 0.079 | 1086.3 | Relay4 (backup) |  |

**Table 9.** Relay's operating time for three phase fault on different buses in presence of DG1 and UFCL

| Interval (s) | Performance Time (s) | Relay Flow (A) | Number of Relay (Main and Backup) | Fault Location |
|---|---|---|---|---|
| 0.574 | 0.42 | 988.1 | Relay2 (Main) | Bus 3 |
|  | 0.984 | 980.3 | Relay1 (backup) |  |
| 0.348 | 0.21 | 676.5 | Relay3 (Main) | Bus 4 |
|  | 0.558 | 676.5 | Relay2 (backup) |  |
| 0.301 | 0.024 | 311.63 | Relay6 (Main) | Bus 6 |
|  | 0.335 | 251.2 | Relay4 (backup) |  |
| 0.365 | 0.444 | 1086.3 | Relay5 (Main) | DG's bus |
|  | 0.079 | 1086.3 | Relay4 (backup) |  |

The proposed method using UFCL suitably limits the short-circuit current flowing through the overcurrent relay during faults, so delay in operating overcurrent relay achieve coordination with the LVRT operation of the DGs in that way, the DGs unit will be allowed to disconnect first (after remaining preventing unintentional islanding.

## 5. Conclusions

In this paper, it was found that using distributed generation units in the distribution grids is increasing short-circuit level and it's caused to disrupt intervals in the coordination of overcurrent relays. Using unidirectional fault current limiters which is located in the tie-feeder between downstream and upstream grid and during different scenarios for all types of DGs, it was found that they return the performance time of main and backup relays to the permissible range (0.3-0.6) second. UFCL has low resistance in the normal operation as well as short-circuit in the downstream grid and, has high resistance in the short-circuit of upstream grid.

**References**

[1] H. Haggi, F. Hasanzad, and M. Golkar, "Security-Constrained Unit Commitment considering large-scale compressed air energy storage (CAES) integrated with wind power generation," *International Journal of Smart Electrical Engineering,* vol. 6, pp. 127-134, 2017.

[2] B. J. Brearley and R. R. Prabu, "A review on issues and approaches for microgrid protection," *Renewable and Sustainable Energy Reviews,* vol. 67, pp. 988-997, 2017.

[3] P. T. Manditereza and R. Bansal, "Renewable distributed generation: The hidden challenges–A review from the protection perspective," *Renewable and Sustainable Energy Reviews,* vol. 58, pp. 1457-1465, 2016.

[4] M. Norshahrani, H. Mokhlis, A. Bakar, A. Halim, J. J. Jamian, and S. Sukumar, "Progress on protection strategies to mitigate the impact of renewable distributed generation on distribution systems," *Energies,* vol. 10, p. 1864, 2017.

[5] F. Rahmani, M. A. Robinson, and M. Barzegaran, "Cool roof coating impact on roof-mounted photovoltaic solar modules at texas green power microgrid," *International Journal of Electrical Power & Energy Systems,* vol. 130, p. 106932, 2021.

[6] M. Ansari, M. Zadsar, S. S. Sebtahmadi, and M. Ansari, "Optimal sizing of supporting facilities for a wind farm considering natural gas and electricity networks and markets constraints," *International Journal of Electrical Power & Energy Systems,* vol. 118, p. 105816, 2020.

[7] V. Telukunta, J. Pradhan, A. Agrawal, M. Singh, and S. G. Srivani, "Protection challenges under bulk penetration of renewable energy resources in power systems: A review," *CSEE journal of power and energy systems,* vol. 3, pp. 365-379, 2017.

[8] S. Katyara, L. Staszewski, and Z. Leonowicz, "Protection coordination of properly sized and placed distributed generations–methods, applications and future scope," *Energies,* vol. 11, p. 2672, 2018.

[9] H. C. Kiliçkiran, İ. Şengör, H. Akdemir, B. Kekezoğlu, O. Erdinç, and N. G. Paterakis, "Power system protection with digital overcurrent relays: A review of non-standard characteristics," *Electric Power Systems Research,* vol. 164, pp. 89-102, 2018.

[10] A. J. Urdaneta, L. G. Pérez, and H. Restrepo, "Optimal coordination of directional overcurrent relays considering dynamic changes in the network topology," *IEEE Transactions on Power Delivery,* vol. 12, pp. 1458-1464, 1997.

[11] A. Urdaneta, L. Perez, J. Gomez, B. Feijoo, and M. Gonzalez, "Presolve analysis and interior point solutions of the linear programming coordination problem of directional overcurrent relays," *International Journal of Electrical Power & Energy Systems,* vol. 23, pp. 819-825, 2001.

[12] F. Jiang, S. Cheng, C. Tu, Q. Guo, R. Zhu, X. Li*, et al.*, "Design consideration of a dual-functional bridge-type fault current limiter," *IEEE Journal of Emerging and*

*Selected Topics in Power Electronics,* vol. 8, pp. 3825-3834, 2019.

[13] F. Coffele, C. Booth, and A. Dyśko, "An adaptive overcurrent protection scheme for distribution networks," *IEEE Transactions on Power Delivery,* vol. 30, pp. 561-568, 2014.

[14] A. A. Memon and K. Kauhaniemi, "A critical review of AC Microgrid protection issues and available solutions," *Electric Power Systems Research,* vol. 129, pp. 23-31, 2015.

[15] J. Keller and B. Kroposki, "Understanding fault characteristics of inverter-based distributed energy resources," National Renewable Energy Lab.(NREL), Golden, CO (United States)2010.

[16] S. Australia, "Grid connection of energy systems via inverters," *Part 3: Grid protection requirements,* p. 7, 2005.

[17] R. Teodorescu, M. Liserre, and P. Rodriguez, *Grid converters for photovoltaic and wind power systems* vol. 29: John Wiley & Sons, 2011.

[18] N. K. Roy and H. R. Pota, "Current status and issues of concern for the integration of distributed generation into electricity networks," *IEEE Systems journal,* vol. 9, pp. 933-944, 2014.

[19] G. Antonova, M. Nardi, A. Scott, and M. Pesin, "Distributed generation and its impact on power grids and microgrids protection," in *2012 65th Annual Conference for Protective Relay Engineers*, 2012, pp. 152-161.

[20] C. Rahmann, H.-J. Haubrich, A. Moser, R. Palma-Behnke, L. Vargas, and M. Salles, "Justified fault-ride-through requirements for wind turbines in power systems," *IEEE Transactions on Power Systems,* vol. 26, pp. 1555-1563, 2011.

[21] T. Agarwal, P. Niknejad, F. Rahmani, M. Barzegaran, and L. Vanfretti, "A time-sensitive networking-enabled synchronized three-phase and phasor measurement-based monitoring system for microgrids," *IET Cyber-Physical Systems: Theory & Applications,* 2021.

[22] M. G. M. Zanjani, K. Mazlumi, and I. Kamwa, "Application of μPMUs for adaptive protection of overcurrent relays in microgrids," *IET Generation, Transmission & Distribution,* vol. 12, pp. 4061-4068, 2018.

[23] H. Haggi, W. Sun, and J. Qi, "Multi-Objective PMU Allocation for Resilient Power System Monitoring," in *2020 IEEE Power & Energy Society General Meeting (PESGM)*, 2020, pp. 1-5.

[24] H. Yazdanpanahi, Y. W. Li, and W. Xu, "A new control strategy to mitigate the impact of inverter-based DGs on protection system," *IEEE Transactions on Smart grid,* vol. 3, pp. 1427-1436, 2012.

[25] R. K. Varma, S. A. Rahman, V. Atodaria, S. Mohan, and T. Vanderheide, "Technique for fast detection of short circuit current in PV distributed generator," *IEEE Power and Energy Technology Systems Journal,* vol. 3, pp. 155-165, 2016.

[26] T.-N. Preda, K. Uhlen, and D. E. Nordgård, "An overview of the present grid codes for integration of distributed generation," 2012.

[27] H.-C. Jo, S.-K. Joo, and K. Lee, "Optimal placement of superconducting fault current limiters (SFCLs) for protection of an electric power system with distributed generations (DGs)," *IEEE Transactions on Applied Superconductivity,* vol. 23, pp. 5600304-5600304, 2012.

[28] S. B. Naderi, M. Negnevitsky, A. Jalilian, M. T. Hagh, and K. M. Muttaqi, "Low voltage ride-through enhancement of DFIG-based wind turbine using DC link switchable resistive type fault current limiter," *International Journal of Electrical Power & Energy Systems,* vol. 86, pp. 104-119, 2017.

[29] M. Ansari, M. Ansari, J. Valinejad, and A. Asrari, "Optimal daily operation in smart grids using decentralized bi-level optimization considering unbalanced optimal power flow," in *2020 IEEE Texas Power and Energy Conference (TPEC)*, 2020, pp. 1-6.

[30] F. Van Overbeeke, "Fault current source to ensure the fault level in inverter-dominated networks," in *Proc. CIRED*, 2009, pp. 1-4.

[31] R. Akella, H. Tang, and B. M. McMillin, "Analysis of information flow security in cyber–physical systems," *International Journal of Critical Infrastructure Protection,* vol. 3, pp. 157-173, 2010.

[32] O. Badran, S. Mekhilef, H. Mokhlis, and W. Dahalan, "Optimal reconfiguration of distribution system connected with distributed generations: A review of different methodologies," *Renewable and Sustainable Energy Reviews,* vol. 73, pp. 854-867, 2017.

[33] P. Piya, M. Ebrahimi, M. Karimi-Ghartemani, and S. A. Khajehoddin, "Fault ride-through capability of voltage-controlled inverters," *IEEE Transactions on Industrial Electronics,* vol. 65, pp. 7933-7943, 2018.

[34] S. Beheshtaein, R. Cuzner, M. Savaghebi, and J. M. Guerrero, "Review on microgrids protection," *IET Generation, Transmission & Distribution,* vol. 13, pp. 743-759, 2019.

[35] M. N. Alam, "Adaptive protection coordination scheme using numerical directional overcurrent relays," *IEEE Transactions on Industrial Informatics,* vol. 15, pp. 64-73, 2018.

[36] S. Chaitusaney and A. Yokoyama, "Prevention of reliability degradation from recloser–fuse miscoordination due to distributed generation," *IEEE Transactions on Power Delivery,* vol. 23, pp. 2545-2554, 2008.

[37] M. Hayerikhiyavi and A. Dimitrovski, "Gyrator-Capacitor Modeling of a Continuously Variable Series Reactor in Different Operating Modes," *arXiv preprint arXiv:2104.07184,* 2021.

[38] M. Hayerikhiyavi and A. Dimitrovski, "Comprehensive Analysis of Continuously Variable Series Reactor Using GC Framework," *arXiv preprint arXiv:2103.11136,* 2021.

[39] J. Kennedy, P. Ciufo, and A. Agalgaonkar, "A review of protection systems for distribution networks embedded with renewable generation," *Renewable and Sustainable Energy Reviews,* vol. 58, pp. 1308-1317, 2016.